\documentclass[prl,twocolumn,superscriptaddress,longbibliography]{revtex4-2}
\usepackage[utf8]{inputenc}
\usepackage[american]{babel}
\usepackage[T1]{fontenc}
\usepackage[pdftex]{graphicx}  
\usepackage{xcolor}
\usepackage{dcolumn}
\usepackage{bm}
\usepackage{dsfont}
\usepackage{amsmath,amsthm,amssymb}
\usepackage{braket}
\usepackage{wrapfig}
\usepackage{mathtools}
\usepackage{float}
\usepackage{physics}
\usepackage[normalem]{ulem}

\usepackage{hyperref}
\hypersetup{colorlinks,bookmarksopen,bookmarksnumbered,citecolor=red,linkcolor=red,pdfstartview=false,urlcolor=red}

\begin{document}
\title{Enhancing Revivals Via Projective Measurements in a Quantum Scarred System}

\author{Alessio Paviglianiti}
\affiliation{International School for Advanced Studies (SISSA), via Bonomea 265, 34136 Trieste, Italy}
\author{Alessandro Silva}
\affiliation{International School for Advanced Studies (SISSA), via Bonomea 265, 34136 Trieste, Italy}

\begin{abstract}
    Quantum many-body scarred systems exhibit atypical dynamical behavior, evading thermalization and featuring periodic state revivals. In this Letter, we investigate the impact of projective measurements on the dynamics in the scar subspace for the paradigmatic PXP model, revealing that they can either disrupt or enhance the revivals. Local measurements performed at random times rapidly erase the system's memory of its initial conditions, leading to fast steady-state relaxation. In contrast, a periodic monitoring amplifies recurrences and preserves the coherent dynamics over extended timescales. We identify a measurement-induced phase resynchronization, countering the natural dephasing of quantum scars, as the key mechanism underlying this phenomenon. 
\end{abstract}
\maketitle

Understanding the out-of-equilibrium properties of quantum many-body systems is an important quest in condensed matter and statistical physics~\cite{polkovnikov2011,eisert2015,heyl2018}, and its significance is growing rapidly with the advent of quantum technologies. Studies of quench dynamics~\cite{kinoshita2006,calabrese2006,alba2018,mitra2018} have led to significant advances in fundamental problems, such as quantum thermalization~\cite{deutsch1991,langen2015,ueda2020,rigol2007,rigol2008,dalessio2016,gogolin2016}, information propagation in many-body systems~\cite{richerme2014,swingle2018,xu2019,lewisswan2019}, and integrability~\cite{caux2013,bertini2016,vidmar2016,castroalvaredo2016,alba2017,denardis2018}. The class of relevant problems was later extended to quantum circuits~\cite{brown2010,nahum2017,bertini2019,fisher2023}, providing new insights into complexity and random quantum states. More recently, the inclusion of measurements in the dynamics has led to the discovery of 
intriguing non-equilibrium phenomena like measurement-induced phase transitions (MIPTs)~\cite{li2018,skinner2019,gullans2020,choi2020,ippoliti2021,coppola2022,paviglianiti2023,difresco2024,lirasolanilla2024,turkeshi2021,turkeshi2022,block2022,turkeshi2024,li2024}, and fostered the exploration of non-unitary quantum state evolution~\cite{murciano2023,garratt2023,paviglianiti2024,turkeshi2023,legal2023,paviglianiti2024_3,zerba2023}. Since their introduction, MIPTs have been explored in a wide class of Hamiltonian and circuit models, revealing a rich phase classification characterized not only by entanglement, but also by quantum fluctuations~\cite{agrawal2022,barratt2022,oshima2023} and complexity~\cite{lumia2024,bejan2024,fux2024,paviglianiti2024_2}.

In monitored dynamics, the choice of the initial state is typically irrelevant, as measurements gradually erase the memory of the starting conditions. An interesting yet still unexplored question is whether this mechanism holds in localized~\cite{nandkishore2014,altman2015,laflorencie2016} or quantum scarred systems~\cite{schecter2019,serbyn2020,surace2021,surace2023,logaric2024}, which are known to elude thermalization in the absence of measurements. In this Letter, we explore the impact of monitoring in the PXP spin chain~\cite{turner2017,turner2018,ho2019,turner2021,desaules2022,lin2019,giudici2024,yuan2022}, an effective model for Rydberg atom arrays, where certain initial states, such as the Néel state, exhibit long-time revivals~\cite{bernien2017,choi2019,hudomal2022}. We first analyze the PXP dynamics with local projective measurements applied at random times, comparing the evolution of uniform and Néel initial states. While eventually they both approach a common steady state, they retain distinct dynamical features, such as entanglement growth rate. Next, we focus on the Néel state under periodic monitoring synchronized with its revivals, demonstrating a measurement-induced fidelity enhancement relative to the unmeasured evolution. Surprisingly, although the measurements are local, their effects propagate over long distances, which we attribute to the large multipartite entanglement generated by the PXP dynamics. Finally, we approach this revival enhancement from the perspective of many-body scars, revealing that measurements resynchronize quantum scarred eigenstates, counteracting their natural dephasing.

\textit{Model --} Let us now introduce the PXP model and the measurements we consider. Throughout this Letter, we denote the down and up eigenstates of $\hat{\sigma}^z$ by $\ket{\circ}$ and $\ket{\bullet}$, respectively, following commonly used notation.

The PXP model~\cite{turner2017,turner2018} is defined by
\begin{equation}\label{hamiltonian}
    \hat{H} = \sum_{j=1}^N \hat{P}_{j-1}\hat{\sigma}^x_j\hat{P}_{j+1},
\end{equation}
where $\hat{P}_j = \left(\hat{\mathds{1}}-\hat{\sigma}^z_j\right)/2 = \ket{\circ}_j\bra{\circ}_j$ is the projector onto the down state, and either open or periodic boundary conditions (OBC or PBC, respectively) are assumed~\footnote{For OBC, the boundary terms take the form $\hat{\sigma}_1^x\hat{P}_2+\hat{P}_{N-1}\hat{\sigma}_N^x$.}. This Hamiltonian allows for a spin flip on a given site only if its neighbors contain no excitations $\ket{\bullet}$. As a consequence, assuming that the system is initially prepared in a state that does not contain pairs of neighboring excitations of type $\ket{\bullet \bullet}$, these cannot be generated by the dynamics. In the following, we consider two choices of initial states that satisfy this requirement, namely, the uniform state $\ket{\Psi_0^\text{unif}} = \otimes_{j=1}^N\ket{\circ}_j$ and the Néel state $\ket{\Psi_0^\text{Néel}} = \otimes_{j=1}^{N/2} \ket{\bullet}_{2j-1}\ket{\circ}_{2j}$. In addition, for the monitoring we consider local measurements of the operator $\hat{\sigma}_j^z$, which are unable to create adjacent excitations. As a result, the system only explores a portion of the full Hilbert space called Fibonacci cube (for OBC) or Lucas cube (for PBC), spanned by all computational basis states without neighboring excitations~\cite{turner2017,turner2018}.

A well known property of the non-integrable model of Eq.~\eqref{hamiltonian} is the presence of quantum scars. In detail, the spectrum hosts $N$ eigenstates with unexpected properties, such as low entanglement entropy, large multipartite entanglement, high overlap with the Néel state, and violation of the Eigenstate Thermalization Hypothesis (ETH). These scarred states are responsible for the anomalous dynamics of $\ket{\Psi_0^\text{Néel}}$, which shows imperfect periodic revivals at times $t=nT$, $n\in\mathds{N}$, where $T\approx 4.72$ is the period~\cite{turner2018}. This is explained by considering that scars are approximately equally spaced in energy with a separation $\Delta E\approx 2\pi/T \approx 1.33$, and thus they get approximately in phase every multiple of $T$. As we show later, measurements can be used to reinforce this rephasing, countering the inhomogeneous dephasing due to the slightly anharmonic scar energy spectrum.

\textit{Dynamics with random measurements --} We investigate the evolution of the uniform and Néel states under the PXP Hamiltonian interspersed with projective measurements of $\hat{\sigma}_j^z$. Each site is monitored independently and at random times, assuming a constant probability per unit time $\gamma$. This is the standard framework considered for measurement-induced criticality, where the parameter $\gamma$ tunes a phase transition in the long-time entanglement attained by the system.
\begin{figure}[t]
    \centering
    \includegraphics[width=\columnwidth]{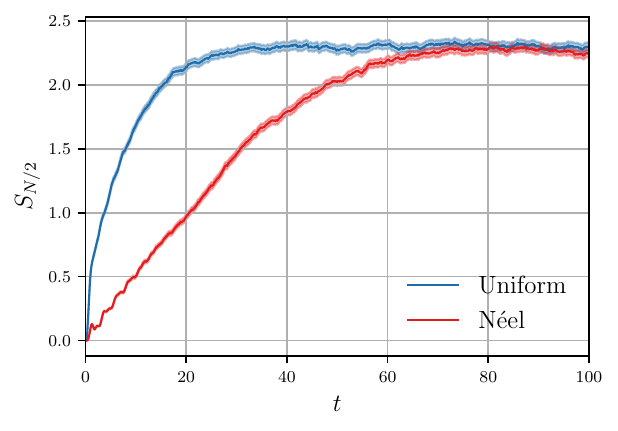}
    \caption{Half-chain entanglement entropy dynamics for the two initial states, using $N=28$, $\gamma=0.04$, and assuming OBC. Data is averaged over $500$ random realizations. The entropy growth rate depends on the initial conditions, but the saturation value is unique. Measurements enhance the amplitude of revivals, while at the same time reducing the entanglement built up by the system.}
    \label{fig:random_meas}
\end{figure}

The dynamics of entanglement in the presence of measurements depends crucially on the initial state. In Fig.~\ref{fig:random_meas} we present our numerical results for the half-chain entanglement entropy averaged over multiple random realizations of the measurement protocol, denoted by $S_{N/2}$. Simulations are performed with matrix-product states using OBC, implementing a standard TEBD algorithm~\cite{vidal2004,schollwock2011}. The non-thermalizing nature of the Néel state is manifested by a slower entanglement growth compared to the uniform state, which reaches the steady state quicker. Notice however that the stationary state is common in both cases, indicating that measurements indeed erase memory of the initial conditions. In fact, the slow dynamics of the Néel state is a consequence of scars, which constitute a significant portion of $\ket{\Psi_0^\text{Néel}}$; throughout the dynamics, measurements change the overlap of the state with the scar subspace, and since the ETH eigenstates exponentially outnumber the scars it is extremely likely that the monitoring deteriorates this overlap in a finite time. Hence, the saturation observed starting from the Néel initial condition can be understood as a gradual transfer of weight from the scar to the ETH subspaces. This effect can be also visualized by observing that a scar-destroying measurement increases the entanglement velocity $d S_{N/2}/dt$, in agreement with the state becoming more ETH-dominated (see Supplemental Material~\cite{suppl_mat}). We will also show explicitly that performing measurements at random times indeed destroys the scar overlap.

We also observe a MIPT by investigating the scaling properties of the saturation value $S_{N/2}(t\to\infty)$. We detect a volume law phase for $\gamma<\gamma_c=0.013\pm0.002$, whereas the system shows area law behavior at larger rates. This value of $\gamma_c$ was estimated by performing a finite-size scaling analysis~\cite{skinner2019,modak2021,sharma2022,harada2011} to test whether the volume phase is stable or just a finite-size effect, and we observe that the optimal data collapse is consistent with a non-zero critical rate~\cite{suppl_mat}. The low value of $\gamma_c$ as compared to other monitored Hamiltonian models is due to the local PXP constraint forbidding neighboring excitations. Whenever a measurement with outcome $\ket{\bullet}_j$ occurs, the adjacent spins must necessarily be projected to $\ket{\circ}_{j\pm 1}$, thus disentangling $3$ qubits instead of a single one. As a consequence, measurements in the PXP model disrupt entanglement much more effectively than in other systems, and thus the volume phase appears only at very low values of $\gamma$.

\textit{Measurement-enhanced revival dynamics --} Instead of measuring at random times, let us now consider the impact of a periodic monitoring on the dynamics. The approximate revivals of the Néel state make it periodically return to an almost computational-basis state. As we show below, local measurements can be used to correct the imperfect fidelity and strengthen the revivals.
\begin{figure}[t]
    \centering
    \includegraphics[width=\columnwidth]{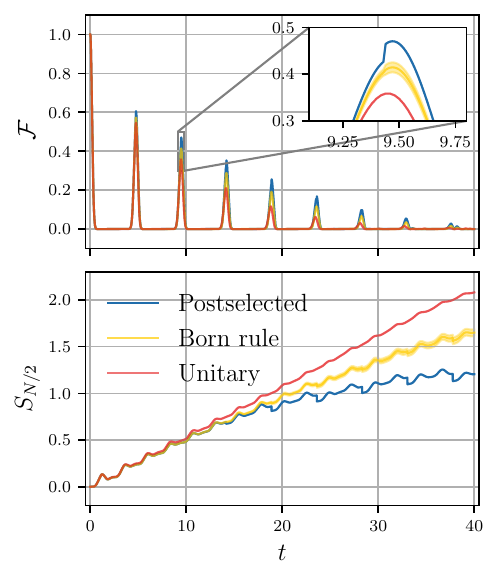}
    \caption{Dynamics of (a) fidelity $\mathcal{F}$ and (b) half-chain entanglement entropy $S_{N/2}$ under a periodic monitoring of the left-most site, starting from the Néel state and using $N=32$. We compare the unmeasured unitary case to both Born rule measurements (for which we average over $200$ random realizations) and postselected $\ket{\bullet}_1$ projections.}
    \label{fig:revival_enhancement}
\end{figure}

Starting from the initial state $\ket{\Psi_0^\text{Néel}}$, we implement the periodically monitored dynamics by alternating a unitary evolution for a duration $T$ and a single projective measurement performed on the boundary spin with $j=1$, assuming OBC. We consider Born rule measurements, where the outcome is sampled randomly, as well as postselected projections to $\ket{\bullet}_1$, which is the state that would be recovered if the revival was perfect. We compute the fidelity
\begin{equation}\label{fid}
    \mathcal{F}(t) = |\braket{\Psi_t}{\Psi_0^\text{Néel}}|^2,
\end{equation}
where $\ket{\Psi_t}$ is the evolved state at time $t$ under the hybrid dynamics. Our numerical results, again obtained with TEBD simulations, are shown in Fig.~\ref{fig:revival_enhancement}a. As expected, the postselected measurements strengthen the revivals, as they explicitly project the boundary spin to a local state compatible with $\ket{\Psi_0^\text{Néel}}$. Surprisingly an enhancement can still be observed even for Born rule measurements, where the first spin has a chance to be projected to $\ket{\circ}_1$, instantly deleting the overlap with the Néel state. This shows that measurements counter the revival decay, extending the duration of coherent behavior shown by the system. We point out that the probability of the postselected trajectory decays exponentially yet slowly in the number of measurements, with each postselected outcome having a probability $\gtrsim 90 \%$ in the time window considered; hence, it can be accessed in realistic implementations.

It is interesting to also analyze the entanglement growth of the periodically monitored dynamics. In absence of measurements, entanglement shows oscillatory behavior but an overall linear increase. Not surprisingly, measurements reduce the overall entanglement, as we observe in Fig.~\ref{fig:revival_enhancement}b for $t\gtrsim 10$. We point out that the three curves coincide at shorter times because we are evaluating the half-chain entanglement entropy, and thus the effects of a measurement on the boundary take time to propagate through a distance $N/2$. At longer times, we observe that the entropy shows sudden jumps, indicating that the measurement on the first site instantaneously impacts the bulk. In other words, despite the measurement being local, it causes a macroscopic collapse of the state that causes long-range effects at extensive distances $\sim N$. This is an explicit manifestation of the large multipartite entanglement~\cite{toth2012,hyllus2012,amico2008} generated by the PXP evolution, which was highlighted in a previous study~\cite{desaules2022}. In detail, the system develops long-range correlations that mediate a non-local action of the measurements, similar to how measuring locally a GHZ state $(\otimes_{j=1}^N\ket{\circ}_j+\otimes_{j=1}^N\ket{\bullet}_j)/\sqrt{2}$ affects all qubits instantly. This also explains why measuring even a single, boundary spin can result in an exponential enhancement of the fidelity.

Besides the boundary qubit monitoring, we also considered measurements of bulk spins and on multiple sites at the same time. In the postselected case, the results are qualitatively similar to those presented in Fig.~\ref{fig:revival_enhancement}, with the revival enhancement becoming stronger the more qubits are measured. In contrast, in the Born rule case this effect gets weaker with the number of monitored spins.

\textit{Quantum scar resynchronization --} Quantum scars constitute the fundamental ingredient for fidelity revivals. We now explore how measurements impact the scar superposition of the Néel state. Not only we observe that periodic measurements leave the overall overlap with the scar subspace almost unchanged, but we highlight that the key mechanism underlying the revival enhancement is a measurement-induced realignment of the scar phases.

\begin{figure}[t]
    \centering
    \includegraphics[width=\columnwidth]{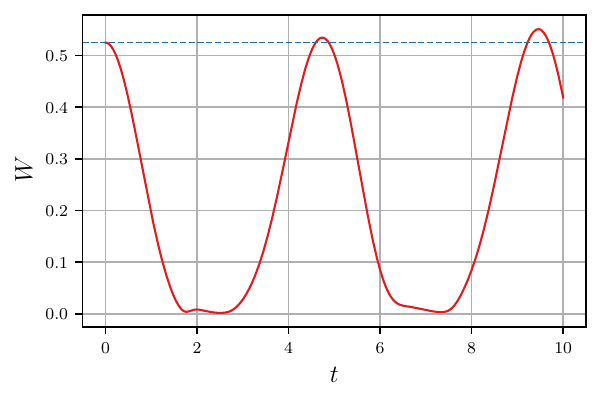}
    \caption{Post-measurement scar subspace weight $W$ as a function of the measurement time $t$, using $N=26$ and PBC. Measurements performed at times $t=nT$ leave the weight almost unchanged, whereas they deteriorate the scar overlap when acting at other times.}
    \label{fig:scar_weight}
\end{figure}

Let us denote the scar states by $\ket{\psi_s}$, $s=1,\dots N$. These can be obtained through exact diagonalization, as described in Ref.~\cite{turner2018}, by looking for the energy eigenstates with lowest entanglement entropy. In the following, we assume PBC. We are interested in understanding how projective measurements modify the overlap of the evolved state $\ket{\Psi_t}=e^{-i\hat{H}t}\ket{\Psi_0^\text{Néel}}$ with the scar subspace. Of course, such an overlap is left invariant by the unitary evolution, and only measurements can change it. We thus assume to project $\ket{\Psi_t}$ locally to the state $\ket{\bullet}_{1}$, corresponding to a postselected measurement~\footnote{Notice that choosing any odd site provides equivalent results, since we are using PBC.}, yielding the post-measurement state
\begin{equation}
    \ket{\Psi^\text{proj}_t} = \frac{(\hat{\mathds{1}}-\hat{P}_1)\ket{\Psi_t}}{\sqrt{1-\bra{\Psi_t}\hat{P}_1\ket{\Psi_t}}}.
\end{equation}
We then introduce the scar weight
\begin{equation}
    W = \sum_{s=1}^N \left|\braket{\psi_s}{\Psi_t^\text{proj}}\right|^2,
\end{equation}
which measures the overlap with the scar subspace after performing a measurement at time $t$. Figure~\ref{fig:scar_weight} shows the time-dependence of $W$. As expected, at almost all times the projection decreases the weight, bringing the state toward the ETH part of the spectrum. However, we observe that by finely tuning the measurement time to match multiples of the period $T$, the measurement leaves $W$ practically unaffected, if not slightly increased. This result is surprising, because no individual scar $\ket{\psi_s}$ is an eigenstate of the projector $\hat{\mathds{1}}-\hat{P}_1$. The specific superposition structure of the state causes the measurement to reshuffle scars without changing their overall weight.

To conclude, we analyze more in detail how measurements performed at $t=T$ modify the individual coefficients of scars appearing in the state superposition. Any arbitrary state $\ket{\Psi}$ can be written as
\begin{equation}\label{state_split}
    \ket{\Psi} = \sqrt{1-W} \ket{\Psi_{ETH}} + \sum_{s=1}^N a_s e^{i \varphi_s} \ket{\psi_s},
\end{equation}
highlighting the scar and ETH contributions. Here $a_s>0$ satisfy $\sum_{s=1}^N a_s^2 = W$, where $W$ is the total scar weight. Since each $\ket{\psi_s}$ is defined up to a phase, we conventionally set it such that the Néel state has $\varphi_s=0$. In order to understand how measurements change the phases, we first consider the state $\ket{\Psi_T} = e^{-i \hat{H}T}\ket{\Psi_0^\text{Néel}}$ evolved by one period, which clearly has $\varphi_s = E_s T$, where $E_s$ are the scar energies. We then perform projective measurements on a compact set of $n$ neighboring qubits (i.e., from $j=1$ to $j=n$), postselecting the outcomes to be $\ket{\bullet}_{2j-1}$ for odd sites and $\ket{\circ}_{2j}$ for even ones, such that they gradually impose the Néel ordering. Finally, we evaluate the post-measurement phases $\varphi_s$.
\begin{figure}[t]
    \centering
    \includegraphics[width=\columnwidth]{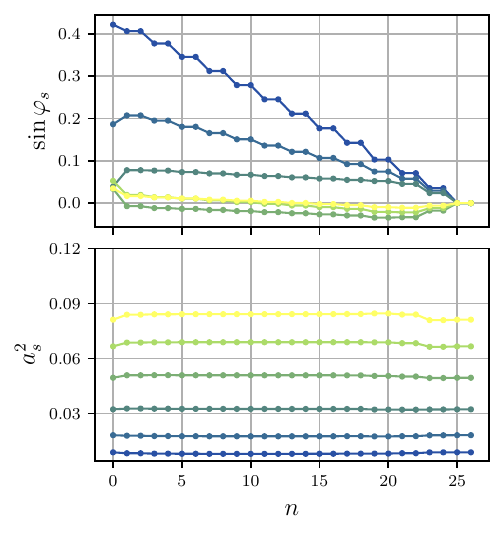}
    \caption{Changes of (a) $\sin \varphi_s$ and (b) $a_s$ as functions of the number $n$ of measured sites on the state $\ket{\Psi_T}$, using $N=26$. $n=0$ corresponds to the unmeasured state, for which $\varphi_s=E_s T$. Different colors correspond to different scar indices $s$, and only scars with $a_s^2|_{n=0}>0.005$ and $\sin \varphi_s |_{n=0} > 0$ are shown (the phases appear in pairs $\pm\varphi_s$ due to the spectrum $E_s$ being symmetric around zero).}
    \label{fig:phases}
\end{figure}

The results of this calculation are presented in Fig.~\ref{fig:phases}a. First, the unmeasured state ($n=0$) manifests different phases for the various scars, which occurs because the energy spacings of $E_s$ are not perfectly homogeneous. Measuring more and more sites mitigates this dephasing, bringing $\varphi_s$ back to zero as they are for the Néel state. In the process, we observe in Fig.~\ref{fig:phases}b that the individual amplitudes $a^2_s$, not only their sum $W$, remain approximately constant. We thus conclude that postselected measurements at $t=T$ implement a rephasing mechanism that realigns scar states without significantly affecting their weights, thus countering the dephasing responsible for the revival decay.

\textit{Conclusions --} We investigated the dynamics of the PXP model under local monitoring, highlighting that measurements can either destroy coherent revivals or enhance them depending on how they are implemented. Despite the non-ergodic nature of the Néel state evolution in the unitary case, measurements performed at random times erase memory of the initial conditions leading the system to an ETH-dominated steady state. Nevertheless, signatures of the scar dynamics are visible in the transient regime, where the Néel state takes longer to relax as compared to the uniform one. In contrast, periodic measurements synchronized with the Néel state revivals enhance the coherent scar dynamics rather than destroying it. While this effect is expected in the case of postselected measurements, we surprisingly observe that it persists also in the Born rule case. We further observe that local measurements cause non-local effects mediated by large multipartite correlations, allowing a single-site monitoring to impact the full system entirely. Finally, we uncover that the periodic monitoring unexpectedly acts as a rephasing process that counters the natural scar dephasing.

Our findings shed light on the physics of monitored non-thermalizing models, which are yet widely unexplored, and on memory loss of initial conditions in MIPTs. It would be interesting to investigate whether the different relaxation timescales for the uniform and Néel states can be observed from a Liouvillian approach, possibly being related to a notion of open system scars that decay slowly. Another question left for future studies is whether an external monitoring can help to fully stabilize the state revivals in other scarred models where measurements can be implemented continuously in times, without being constrained to multiples of a period $T$.

\textit{Acknowledgements --} A.S. would like to acknowledge support from PNRR MUR Project ``Superconducting quantum-classical linked computing systems (SuperLink)'', in the frame of QuantERA2 ERANET COFUND in Quantum Technologies, CUP B53C22003320005.

\bibliography{biblio}

\pagebreak
\widetext
\newpage
\begin{center}
\textbf{\large Supplemental Material}
\end{center}

\setcounter{equation}{0}
\setcounter{figure}{0}
\setcounter{section}{0}

\makeatletter
\renewcommand \thesection{S\@arabic\c@section}
\renewcommand{\theequation}{S\arabic{equation}}
\renewcommand{\thefigure}{S\arabic{figure}}

In this Supplemental Material we provide additional results on the monitored PXP dynamics. First, we illustrate how a scar-destroying measurement increases the entanglement velocity $d S_{N/2}/dt$ of the Néel state, in agreement with the property that ETH-dominated states build up entanglement quicker. Then, we present our finite-size scaling analysis of the long-time entanglement entropy, highlighting the presence of a MIPT.

\section{Entanglement velocity increase after a measurement}

\begin{wrapfigure}{r}{0.47\textwidth}
    \centering
    \includegraphics[width=0.45\textwidth]{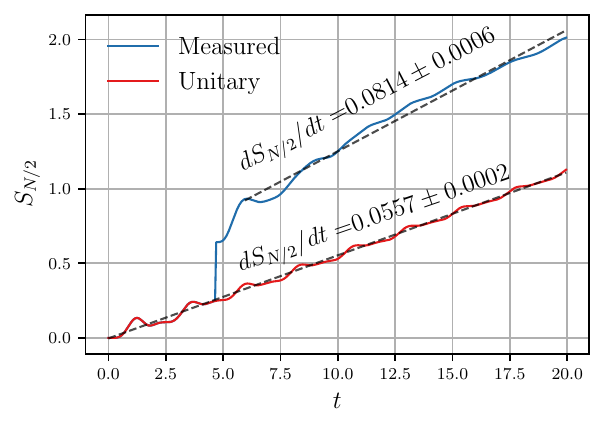}
    \caption{Half-chain entanglement entropy computed for the unitary dynamics (red) and in presence of a single local projection to $\ket{\circ}_{N/2-1}$ applied at $t=T$ (blue). Data for $N=32$ and OBC, starting from the Néel state. The measurement deteriorates the overlap with the scar subspace, and this is reflected by an increased entanglement growth rate (estimated with a linear fit).}
    \label{f:slope_change}
\end{wrapfigure}
The rate of growth of the entanglement entropy $S_{N/2}$ varies depending on the choice of the initial state $\ket{\Psi_0}$~\cite{ho2019}. In detail, the dynamics of the Néel state takes longer to build up entanglement as compared to the case of the uniform initial state. This can be understood by considering Eq.~\eqref{state_split}: the Néel state has a significant weight over the scar subspace, which does not undergo regular scrambling dynamics, but rather evolves in an almost coherent way by displaying approximate revivals.

As argued in the main text, measurements performed at random times generally suppress the weight of the state over the scar subspace. Since the part of the state undergoing scrambling dynamics gains weight, we expect that entanglement will grow faster. To test this, we evolve $\ket{\Psi_0^\text{Néel}}$ with the PXP Hamiltonian and disrupt the dynamics with a single postselected measurement to $\ket{\circ}_{N/2-1}$ performed at time $t=T$. Before the measurement, the approximate Néel revival implies that site $N/2-1$ is close to $\ket{\bullet}_{N/2-1}$ (for even $N/2$); as a consequence, by projecting to $\ket{\circ}_{N/2-1}$ we significantly alter the structure of the state, greatly reducing the scar overlap.

In Fig.~\ref{f:slope_change} we show the entanglement growth after the measurement and we compare it to the case of regular unitary dynamics. As expected, we observe that the measurement produces an increase in the slope $d S_{N/2}/dt$. This further confirms that measurements performed at generic times decrease the scar overlap of the state, thus erasing the coherent revival dynamics and eventually leading the system to a fully scrambled state. 

\section{Finite-size scaling analysis}

\begin{figure}[t]
    \centering
    \begin{minipage}{0.47\linewidth}
        \centering
        \includegraphics[width=\columnwidth]{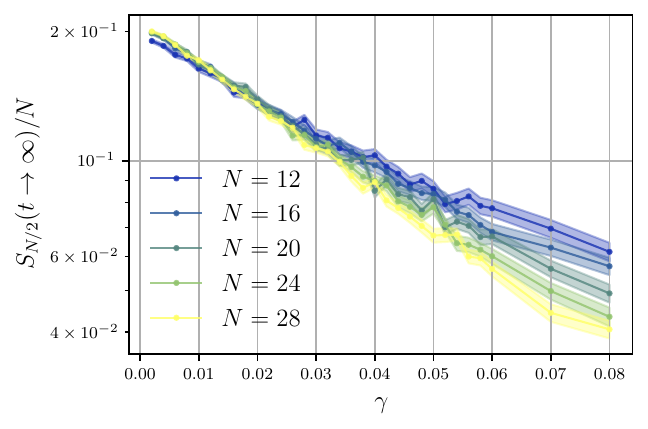}
        \caption{Long-time entropy density $S_{N/2}(t\to\infty)/N$ as a function of the measurement rate $\gamma$, for multiple system sizes $N$. We observe a crossover between a volume law and an area law around $\gamma\approx0.015$.}
        \label{f:entropy_density}
    \end{minipage}
    \hfill
    \begin{minipage}{0.47\linewidth}
        \centering
        \includegraphics[width=\columnwidth]{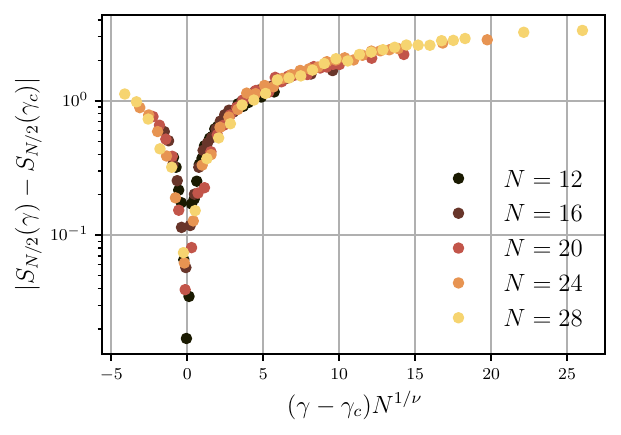}
        \caption{Finite-size scaling analysis of the long-time entanglement entropy, using $N=12,16,20,24,28$. The fitted parameters are $\gamma_c = 0.013\pm0.002$ and $\nu=0.56\pm0.02$, where the uncertainties are estimated by bootstrapping.}
        \label{f:collapse}
    \end{minipage}
\end{figure}

MIPTs are usually characterized in terms of the scaling properties of the trajectory-averaged long-time entanglement entropy $S_{N/2}(t\to\infty)$. In particular, generic interacting models usually exhibit a volume law $S_{N/2}\sim N$ at small measurement rates $\gamma<\gamma_c$, and an area law $S_{N/2}\sim \mathrm{const.}$ for $\gamma>\gamma_c$. In Fig.~\ref{f:entropy_density}, we show the entanglement density $S_{N/2}(t\to\infty)/N$ as a function of the measurement rate $\gamma$. The long-time entropy is evaluated as a long-time average, performed over the time window $t\in[60,80]$. For $\gamma\lesssim 0.015$, the entropy density appears to converge to a constant as $N$ is increased, indicating volume law scaling. In contrast, at large $\gamma$ we notice $S_{N/2}(t\to\infty)/N\to 0$, compatible with the anticipated area law.

At the transition point, the system is expected to manifest scale invariance and universal behavior. To better characterize the transition and estimate the critical point, we performed a finite-size scaling analysis. We use the ansatz
\begin{equation}\label{scaling_ansatz}
    |S_{N/2}(\gamma)-S_{N/2}(\gamma_c)| = f((\gamma-\gamma_c)N^{1/\nu}),
\end{equation}
where $f$ is an unknown universal function, and $\gamma_c$ and $\nu$ are parameters to be optimized. This scaling form proved to be appropriate for other monitored models~\cite{skinner2019,modak2021,sharma2022}. The critical entropies $S_{N/2}(\gamma_c)$ appearing in Eq.~\eqref{scaling_ansatz} are estimated by linear interpolation using data of the closest values of $\gamma$ available. The parameter optimization is performed using a variation of the algorithm proposed in Ref.~\cite{harada2011}, which uses a Bayesian inference technique. The optimal data collapse is shown in Fig.~\ref{f:collapse}. We obtain a finite critical point $\gamma_c=0.013\pm0.002$, indeed suggesting that the MIPT is genuine.

\end{document}